\documentclass[a4paper,12pt]{article}
\pdfoutput=1 
\usepackage{verbatim}

\usepackage{jheppub}

\usepackage[T1]{fontenc}

\title{\boldmath Quarter-BPS $AdS_5$ solutions in M-theory with a $T^2$ bundle over a Riemann surface}

\author{Ibrahima Bah}
\affiliation{Institut de Physique Th\'{e}orique, CEA/ Saclay \\91191 Gif-sur-Yvette Cedex, France}

\emailAdd{Ibrahima.ba@cea.fr}

\abstract{We study and classify quarter-BPS $AdS_5$ systems in M-theory, whose internal six-dimensional geometry is a $T^2$ bundle over a Riemann surface and two interval directions.  The general system presented, provides a unified description of all known $AdS_5$ solutions in M-theory.  These systems are governed by two functions, one that corresponds to the conformal factor of the Riemann surface and another that describes the $T^2$ fibration.  We find solutions that can be organized into two classes.  In the first one, solutions are specified by the conformal factor of the Riemann surface which satisfies a warped generalization of the $SU(\infty)$ Toda equation.  The system in the second class requires the Riemann surface to be $S^2$, $H_2$ or $T^2$.  Class one contains the M-theory $AdS_5$ solutions of Lin, Lunin and Maldacena; the solutions of Maldacena and N\'{u}\~{n}ez; the solutions of Gauntlett, Martelli, Sparks and Waldram; and the eleven-dimensional uplift of the $Y_{p,q}$ metrics.  The second includes the recently found solutions of Beem, Bobev, Wecht and the author.  Within each class there are new solutions that will be studied in a companion paper.}

\preprint{IPhT-T13/054, MCTP-13-12}

\begin{document} 
\maketitle
\flushbottom

\newpage

\section{Introduction}
Five-branes in M-theory are very useful for studying and describing superconformal field theories (SCFT's) in various dimensions.  This story started in pre AdS/CFT \cite{Maldacena:1997re} days with Witten's description \cite{Witten:1997sc} of the strong coupling limit of $\mathcal{N}=2$ quiver gauge theories as a M5-brane wrapping a holomorphic curve of $R^3\times S^1$ in M-theory.  In more recent times, the moduli space of a large class of $\mathcal{N}=2$ quiver gauge theories were studied by extending Witten's descriptions to M5-branes wrapping Riemann surface with defects \cite{Gaiotto:2009hg}.  Building on this, Gaiotto \cite{Gaiotto:2009we} showed that strongly coupled and isolated $\mathcal{N}=2$ SCFT's are classified by M5-branes wrapping punctured Riemann surfaces embedded in a four-dimensional subspace of M-theory.  Constructions from \cite{Gaiotto:2009hg,Gaiotto:2009we} are typically called theories of class $\mathcal{S}$ due to their six-dimensional origin. They do not admit known Lagranian descriptions but a great deal of their physical properties can be deduced from the constructions.  Different $\mathcal{N}=2$ SCFT's are labelled by the genus of the two-dimensional surface and the types of punctures present.  Exactly marginal couplings of $\mathcal{N}=2$ SCFT's correspond to relative positions of punctures, and therefore, the rich S-duality of $\mathcal{N}=2$ SCFT's is described by the various ways of bringing punctures close to each other.  Gaiotto's classification has lead to intense activity which continues to elucidate the properties of $\mathcal{N}=2$ SCFT's; here is an incomplete list of references \cite{Alday:2009aq,Alday:2009qq,Kanno:2009ga,Gaiotto:2010be,Tachikawa:2010vg,Nanopoulos:2010ga,Chacaltana:2010ks,Gaiotto:2011tf,Cecotti:2011rv,Tachikawa:2011yr,Alim:2011ae,Alim:2011kw,Gaiotto:2011xs}.

The wonders of M5-branes continue with the description of a class of three-dimensional $\mathcal{N}=2$ SCFT's \cite{Dimofte:2011ju,Dimofte:2011py,Dimofte:2013iv} from wrapping M5-branes on hyperbolic three-manifolds such as knot complements.  Recently a principle of $c$-extremization, a tool for determining R-symmetry, in two-dimensional $\mathcal{N}=(0,2)$ SCFT's \cite{Benini:2012cz,Benini:2013cda} has been uncovered by studying M5-branes wrapped on four-manifolds.  

Gaiotto's classification was further validated by using AdS/CFT.  In \cite{Gaiotto:2009gz} it was shown that the gravity duals to the $\mathcal{N}=2$ constructions can be described by using Lin, Lunin and Maldacena (LLM) $AdS_5$ geometries \cite{Lin:2004nb} in M-theory\footnote{The LLM system describes the most general half-BPS $AdS_5$ solution in M-theory; it was re-derived in \cite{Gauntlett:2006fk}.  A possible loophole on its generality was plugged in \cite{Colgain:2012kx}.  Nonlinear KK reductions of LLM geometries was presented in \cite{Gauntlett:2008uq}.}.  The main question of interest in this paper is: does Gaiotto's classification of $\mathcal{N}=2$ SCFT's extend to $\mathcal{N}=1$ SCFT's?\footnote{It is important to note that a similar question had been asked in pre AdS/CFT days.  Following Witten's description of $\mathcal{N}=2$ theories using M5-branes, there was an intense and interesting program that tried to describe $\mathcal{N}=1$ field theories by using a M5-brane wrapped on a holomorphic curve in a six dimensional submanifold of M-theory.  Seiberg's description of SQCD \cite{Seiberg:1994bz} was reproduced in \cite{Hori:1997ab} as the dynamics of a M5-brane wrapping a holomorphic curve.  An incomplete list of references is \cite{deBoer:1997zy,deBoer:1998by,deBoer:1998rm,Giveon:1997sn,Giveon:1998sr}.  In more recent times the authors of \cite{Tachikawa:2011ea} described some class $\mathcal{S}$ theories with $\mathcal{N}=1$ supersymmetry by using holomorphic curves.}  Our approach to this question is to study how LLM $AdS_5$ solutions, which  preserve eight supercharges, can be generalized to $AdS_5$ systems in M-theory that preserve four supercharges.  To this end, we classify warped $AdS_5$ systems and solutions in M-theory where the internal six-dimensional manifold is a $T^2$ bundle over a closed Riemann surface and two interval directions.  Next, we provide motivation for this approach by reviewing some of the milestones in addressing this question, and what we have learned from them.  

First we review some field theoretic approaches to $\mathcal{N}=1$ extensions of Gaiotto theories.  The authors of \cite{Maruyoshi:2009uk}, for example, considered mass deformations\footnote{Mass deformation refers to giving masses to chiral adjoints in $\mathcal{N}=2$ vector multiplets.  Weakly coupled vector multiplets are present at various S-dual corners.} of Gaiotto theories that break $\mathcal{N}=2$ to $\mathcal{N}=1$.  It was shown that such deformations lead to infinite classes of new isolated $\mathcal{N}=1$ SCFT's which admit generalized versions of Seiberg dualities inherited from S-duality.  In \cite{Bah:2011je} direct constructions of new $\mathcal{N}=1$ quiver gauge theories from Gaiotto theories were considered.  By using $a$-maximization \cite{Intriligator:2003jj} and various tools for analysing $\mathcal{N}=1$ SCFT's, the necessary conditions for constructing new and isolated $\mathcal{N}=1$ SCFT's from Gaiotto theories were discussed.  These theories do not, generically, describe the IR dynamics of any known $\mathcal{N}=1$ deformations of class $\mathcal{S}$ theories as in \cite{Maruyoshi:2009uk}.  Both works hint at the existence of $\mathcal{N}=1$ structures that generalize Gaiotto's classification.   

In \cite{Benini:2009mz} the IR limit of $\mathcal{N}=2$ SCFT's, describing M5-branes wrapped on genus $g>1$ Riemann surfaces without punctures,  with $\mathcal{N}=1$ mass deformations were studied in detail.  Some time ago, the gravity duals of such configurations of M5-branes were described by Maldacena and N\'{u}\~nez (MN) in \cite{Maldacena:2000mw}.  The holographic RG flows from M5-branes on Riemann surfaces were discussed by using seven-dimensional gauged supergravity \cite{Pernici:1984xx}, which uplifts to M-theory \cite{Cvetic:1999xp,Liu:1999ai}.  The authors of \cite{Maldacena:2000mw} identified two IR $AdS_5$ fixed points, one preserving eight supercharges (MN2) and the other preserving four supercharges (MN1).  The field theory dual to the MN2 geometry is the class $\mathcal{S}$ theory corresponding to the genus $g>1$ Riemann surface without punctures \cite{Gaiotto:2009gz}.  In \cite{Benini:2009mz} it is argued that the field theory dual to the MN1 solution is the mass deformed theory dual to MN2.  This was the first extension of class $\mathcal{S}$ theories to $\mathcal{N}=1$ by using gravity.  

The RG flows of \cite{Maldacena:2000mw} have more than two fixed points.  In \cite{Bah:2011vv,Bah:2012dg} (B$^3$W) the authors describe a one-parameter family of quarter-BPS $AdS_5$ solutions in M-theory that emerge as IR fixed points of a stack of M5-branes wrapping a Riemann surface.  Furthermore, by using the tools in \cite{Bah:2011je}, the field theory duals were constructed by using building blocks in Gaiotto's constructions.  These theories do not emerge from known deformations of $\mathcal{N}=2$ class $\mathcal{S}$ theories.  The difference between the solutions comes from how supersymmetry is preserved by the M5-branes' world-volume theory.  Now we review this aspect of these solutions as they provide principle guidance on how we should extend LLM to $\mathcal{N}=1$.  

When branes of any type are wrapped on curved manifolds, supersymmetry of the world volume theory is broken.  Some of the supercharges can be preserved if the theory is topologically twisted \cite{Witten:1988ze,Bershadsky:1995qy}.  The main problem is that supersymmetry for field theories requires globally defined constant spinors that are associated to the conserved supercharges.  However on curved manifolds this is hard to come by since the supercharges will satisfies a non-trivial differential equation which involves the covariant derivative.  When a field theory is topologically twisted, a background gauge field, valued in the R-symmetry, is turned on and tuned to cancel the contribution of the spin connection in the covariant derivative acting on the supercharges.  This latter condition is equivalent to tuning the fluxes from the gauge fields to cancel the curvature two form.  The number of ways this can be done enumerates the possible supersymmetric configurations.  The MN solutions are obtained by considering topological twists of the field theory living on M5-branes \cite{Maldacena:2000mw}.   

The world volume theory of a stack of $N$ M5-branes is the $A_{N-1}$ $(2,0)$ six-dimensional SCFT.  The theory preserves 16 supercharges and has an $SO(5)$ R-symmetry group.  When we wrap M5-branes on a Riemann surface, we can preserve supersymmetry by considering the possible twists of $(2,0)$ SCFT.  Since the spin connection of a two-dimensional Riemann surface is $SO(2)$-valued, we need to turn on a $U(1)$-valued gauge field from the R-symmetry \cite{Maldacena:2000mw}.  The rank of $SO(5)$ is two, therefore we have two-dimensional space of gauge fields to choose from.  Tuning the sum of fluxes from the two Cartan $U(1)$'s to cancel the curvature allows for a one-parameter family of supersymmetric configurations which generically preserve four supercharges \cite{Maldacena:2000mw,Bah:2011vv,Bah:2012dg}.  From the point of view of the M5-branes in M-theory, the Riemann surface is embedded into a Calabi-Yau three-fold and the local geometry is two line bundles over the Riemann surface.  The vanishing of the first Chern class for the CY3 fixes the sum of the degrees of the line bundles to the curvature of the Riemann surface \cite{Bah:2011vv,Bah:2012dg}.  At the end of the day, the $SO(5)$ R-symmetry group is broken to $U(1)^2$; from M-theory point of view, these $U(1)$'s come from the phases of the line bundles.  One linear combination is a flavor $U(1)$ for the field theory.

The solutions of B$^3$W are warped product of $AdS_5 \times \mathcal{C}_g \times \widetilde{S}^4$ where $\mathcal{C}_g$ is a Riemann surface of genus $g$ and $\widetilde{S}^4$ is a squashed four-sphere with $U(1)^2$ isometry.  The circles are generically fibred over $\mathcal{C}_g$.  These isometries corrrespond to the Cartan $U(1)$'s from the broken $SO(5)$ R-symmetry, and the phases of the line bundles.  The sum of the degrees of the circle fibrations is fixed to $2(g-1)$ by the twist condition.  The main result of B$^3$W is that each supersymmetric configuration from the topological twists flows to an $AdS_5$ geometry.  When one of the line bundles is trivial, i.e. the degree of one of the circle fibration vanishes, the system preserves eight supercharges and the dual field theory has $\mathcal{N}=2$ supersymmetry.  The isometry of $\widetilde{S}^4$ enhances to $U(1)\times SU(2)$ which corresponds to the $\mathcal{N}=2$ R-symmetry.  The solution is MN2 and the field theory is from Gaiotto's constructions.  For this case, the Riemann surface is embedded into a four-dimensional space as it is expected for $\mathcal{N}=2$ class $\mathcal{S}$ theories.  When the degrees of the two fibrations are equal, $\widetilde{S}^4$ also has a $U(1)\times SU(2)$ isometry but the solution preserves four supercharges.  The $SU(2)$ is a flavor symmetry and the solution is MN1.  Modulo the MN solutions, we have a one-parameter family of $AdS_5 \times \mathcal{C}_g$ describing the different ways we can supersymmetrically wrap M5-branes on a Riemann surface.  The dual $\mathcal{N}=1$ SCFT's have a $U(1)$ flavor symmetry in addition to the $U(1)$ R-symmetry.

An important aspect of B$^3$W solutions is that the relative warping between the $AdS_5$ and $\mathcal{C}_g$ is constant.  This reflects the fact that the geometry emerges solely from the wrapped M5-branes whose world-volume in the UV is the Minkowski slice in $AdS_5$ and $\mathcal{C}_g$.  The RG flow can only induce a radially dependent relative warp factor between these two subspaces.  This pictures changes if there are other branes localized on the Riemann surface.  The radial RG coordinate for such branes would be different, and therefore the relative warping between the Minkowski and the Riemann surface will generically depend on other coordinates.  This important feature of these solutions is observed in the description of gravity duals of $\mathcal{N}=2$ SCFT's \cite{Gaiotto:2009gz}.    

Gaiotto and Maldacena (GM) use the $AdS_5$ LLM system in M-theory \cite{Lin:2004nb} to describe gravity duals of $\mathcal{N}=2$ field theories from M5-branes on punctured Riemann surface \cite{Gaiotto:2009gz}.  The internal geometry of LLM is $\mathcal{C}_g \times \widetilde{S}^4$.  The isometries of the internal $\widetilde{S}^4$ is $U(1)\times SU(2)$ corresponding to the R-symmetry of the dual field theories.  The circle is fibred over $\mathcal{C}_g$ with degree one.  Locally, the metric on the Riemann surface is conformally flat.  The conformal factor depends on the coordinates of the Riemann surface and on the interval of the $\widetilde{S}^4$.  It is the single function that determines solutions and it satisfies the $SU(\infty)$ Toda equation.  When the conformal factor is separable, the Toda equation reduces to the Liouville equation for the part that depends on the Riemann surface coordinates.  The two-dimensional geometries obtained from such equation are the constant curvature ones, $\mathbb{H}_2$, $T^2$ and $S^2$.  The only regular solution is the one with $\mathbb{H}_2$, which can be replaced with a genus $g>1$ closed Riemann surface by modding with a Fuchsian subgroup.  This solutions is MN2.

From the point of view of Gravity, adding punctures on $\mathcal{C}_g$ corresponds to adding localized sources on the Riemann surface in MN2 \cite{Gaiotto:2009gz}.  In the probe approximation, these sources are M5-branes extended along $AdS_5\times S^1$ and sitting at a point where the $S^2$ shrinks as to preserve the $\mathcal{N}=2$ R-symmetry \cite{Gaiotto:2009gz}.  When these probes are backreacted, the geometry near a puncture should be an $AdS_7 \times S^4$ throat.  The conformal factor in LLM must interpolate between MN2 to $AdS_7 \times S^4$.  When the Riemann surface admits a translation direction or $U(1)$ isometry, the $SU(\infty)$ Toda equation can be mapped to an axially symmetric three-dimensional electrostatic problem \cite{Ward:1990qt,Lin:2004nb}.  The solutions of this latter problem are completely determined by boundary conditions, moreover they satisfy the superposition principle \cite{Gaiotto:2009gz}.  The MN2 solution and the $AdS_7\times S^4$ solution will correspond to different choices of boundary conditions.  The interpolating solution is trivially obtained by superposition.  GM described how to map choices of punctures to boundary conditions, thereby providing explicit constructions for gravity duals of Gaiotto theories.  

Our goal is to understand how this construction by GM can be generalized to quarter-BPS systems.  The first step is to find the generalization of MN2 solutions, i.e.  all quarter-BPS systems of M5-branes wrapped on a Riemann surface without punctures.  These are exactly the B$^3$W solutions.  The next step is to find the Toda like structure that can describe interpolating solutions between B$^3$W to $AdS_7 \times S^4$.  We expect such system to preserve the same isometries as B$^3$W similar to LLM and MN2.  The internal geometry should be a $T^2$ bundle over $\mathcal{C}_g$ with two intervals.  Naively we expect the conformal factor of the Riemann surface to depend on interval coordinates similar to LLM.  In this paper we classify $AdS_5$ systems of this type in M-theory.  
 
In section \ref{genesys} we review the general conditions for supersymmetric $AdS_5$ solutions in M-theory as described in \cite{Gauntlett:2004zh}.  We reduce the system on the most general ansatz for a $T^2$ bundle over a Riemann surface.  We use the equations to refine the ansatz and find coordinates that trivialize many conditions.  In section \ref{cansys} we summarize the main results of this exercise, i.e. we write the most general metric and the necessary system of equations.  The eager reader can jump to this section and review details later.  The residual equations are not readily solvable.  In section \ref{solve} we discuss cases when we can solve the system of equations.  The general metric for solvable systems is described in section \ref{solmetric}.  In section \ref{classI} we discuss two classes of solutions, one that includes the MN1 solutions and a set of solutions described in \cite{Gauntlett:2004zh}, and another that includes the LLM system.  The solutions in these classes are similar to LLM in that they are governed by a single function corresponding to the conformal factor of the Riemann surface.  This function in both cases satisfies a warped generalization of the $SU(\infty)$ Toda equation.  This equation plays the same role for MN1 as does the Toda equation for MN2 in LLM.  We expect it to interpolate between MN1 to $AdS_7\times S^4$.  The general metric for these classes are (\ref{classIa}) and (\ref{class2metric}), respectively.

In section \ref{classII} we describe a class of solution where the Riemann surface is always one of the constant curvature type.  The left over system of equations are on the interval directions.  There are many more solutions in this class that generalize the B$^3$W solutions.  We present a general formalism for writing them and work out an example that includes B$^3$W.  The general metric for this class is (\ref{classIImetric}).  Finally in section \ref{discussions} we provide a summary of results and discuss the next step in this programme.  The reader is free to jump to this section and return to the body for details.  

The work presented here focuses on understanding the $AdS_5$ systems in M-theory and how to solve them.  In \cite{Bah:2013ta} we perform a more careful study of the solutions found here.  We discuss regularity conditions, compute the central charge for the dual theories and the four-form flux that supports the solutions.  We will also discuss the underlying M5-branes possible punctures on the Riemann surface.  

\newpage
\section{$AdS_5$ in M-theory}\label{genesys}
\subsection{Supersymmetric $AdS_5$ systems in M-theory}\label{GMSWsys}

The necessary and sufficient conditions for supersymmetric $AdS_5$ solutions in M-theory are given in \cite{Gauntlett:2004zh} (GMSW system).  We review this general system and then discuss how we plan to use it.  The metric is
\begin{align}\label{genmetric}
 L^{-4/3} ds^2_{11} &=H^{-1/3} \left[ds^2_{AdS_5}+ \frac{1}{9} \cos^2(\zeta) \left(d\psi + \rho\right)^2 +H ds^2(M_4) + \frac{Hdy^2}{\cos^2(\zeta)} \right].
\end{align} The single length scale of the system, $L$, is factored out.  We will set it to one and turn it on when needed by multiplying the overall metric by $L^{4/3}$.  The metric $ds^2_{AdS_5}$ has unit radius.  Solutions are determined by the four-dimensional space $M_4$.  It corresponds to a one-parameter family of K\"{a}hler metrics with complex structure, $\Omega$, and symplectic structure, $J$, that satisfy
\begin{align}
d_4 \Omega &= \left(i \rho-d_4 \log \left(\cos(\zeta)\right)\right) \wedge \Omega \label{om4}\\
\partial_y \Omega &= \left(-\frac{3}{2y} \tan^2(\zeta) - \partial_y \log \left(\cos(\zeta)\right) \right) \Omega \label{omy}\\
\partial_\psi \Omega &= i \Omega. \label{ompsi}
\end{align} and
\begin{align}
d_4 J&= 0 \label{j4}\\
\partial_{y} J &= -\frac{2}{3} y d_4 \rho  \label{jy}\\
\partial_{\psi} J&= 0 \label{jpsi}
\end{align} where $d_4$ is the exterior derivative on $M_4$.  The single function, $\zeta$, that appears in the metric, depends on $y$ and the coordinates on the K\"{a}hler base, but not on $\psi$.  This latter direction parametrizes a $U(1)$ isometry.  The warp factor $H$ is given by
\begin{equation}
H = \frac{1}{4y^2} \left(1-\cos^2(\zeta) \right). 
\end{equation} 

The four-form flux is given once a solution is fixed 
\begin{equation} \begin{split}
 L^{-2} F_4 &=- \left(\partial_y H \right) \widehat{Vol}_4 +  \sec^2(\zeta) \left(*_4 d_4 H \right) \wedge dy - \frac{1}{9} \cos^4(\zeta) \left(*_4 \partial_y \rho \right) \wedge \left(d\psi+\rho\right) \\
&+  \left[\frac{1}{9} \cos^2(\zeta) *_4 d_4 \rho - \frac{4}{3} H J \right] \wedge dy \wedge \left(d\psi+\rho\right). \end{split} \label{4form}
\end{equation} The system of equations above implies the Bianchi identity and equation of motion:
\begin{equation}
 d\left(*_{11} F_4 \right) = d F_4 =0.
\end{equation} The Hodge star operators on the four-dimensional and eleven-dimensional spaces are $*_4$ and $*_{11}$, respectively.   

Our goal is to classify solutions of this system where the internal geometry contains a $T^2$ bundle on a Riemann surface.  The eleven-dimensional metric already has a $U(1)$ isometry, the $\psi$-circle.  Therefore we can impose one more on the K\"{a}hler base.  In the next subsections, we reduce the system and study the consequences of the $U(1)$ in the base.  Throughout this paper, we will not worry about the flux since it is determined once solutions are known.  We focus solely on finding solutions.  

\subsection{Ansatz for K\"{a}hler base}

We want to impose a $U(1)$ isometry on the base, $M_4$.  At fixed $y$, $M_4$ is K\"{a}hler.  The most general complex four-dimensional metric that also admits a $U(1)$ isometry can be written as
\begin{equation}
ds_4 = e^{2A} \epsilon_1 \bar{\epsilon}_1 + e^{2B} \epsilon_2 \bar{\epsilon}_2 
\end{equation} with
\begin{equation}
\epsilon_1 = d\hat{x}_1 +id\hat{x}_2, \;\;\;\; \epsilon_2 = d\tau + e^{C} (i d\phi+ V).
\end{equation}  The complex vector, $V$, has legs along $\epsilon_1$ and $\bar{\epsilon}_1$ only.  The $\hat{x}_1$ and $\hat{x}_2$ plane coordinatize a Riemann surface that is determined by the conformal factor $e^{2A}$.  The coordinate $\tau$ parametrizes an interval.

The $\phi$ direction is a circle which corresponds to the $U(1)$ isometry, no metric functions depends on it.  In real coordinates, the metric ansatz is
\begin{equation}
ds^2_4 = e^{2A} (d\hat{x}_1^2 +d\hat{x}_2^2) + e^{2B} \left((d\tau + e^{C}V^R)^2 + e^{2C}(d\phi +V^I)^2 \right) \label{ansatz}
\end{equation} where $I$ and $R$ superscripts refer to imaginary and real parts.  It is useful to define the frame fields and volume form
\begin{align}
\eta_\tau &= d\tau + e^{C} V^R \\
\eta_\phi &= d\phi + V^I \\
dR_2 & = d\hat{x}_1 \wedge d\hat{x}_2.
\end{align} The K\"{a}hler and the complex two-forms can be written as
\begin{align} 
J&= e^{2A} dR_2 + e^{2B+C} \eta_\tau \wedge \eta_\phi \\
\Omega &= e^{i(\psi+p \phi)} e^{A+B} \Omega_0 \label{omedef}\\
\Omega_0^R &= d\hat{x}_1 \wedge \eta_\tau - e^C d\hat{x}_2 \wedge \eta_\phi \\
\Omega_0^I &= e^C d\hat{x}_1 \wedge \eta_\phi + d\hat{x}_2 \wedge \eta_\tau.
\end{align}  The parameter $p$ corresponds to the charge of $\Omega$ under the $U(1)$ corresponding to $\phi$.  The charge of $\Omega$ under $\psi$ is fixed by (\ref{ompsi}).   

Any K\"{a}hler metric can be brought to a form where $V^R=0$ and $C=0$ by coordinate transformations.  Since $M_4$ is part of a larger metric, and is K\"{a}hler only at fixed $y$, such transformations will generically turn on $dy$ terms in $\eta_\tau$.  Therefore we cannot turn off $V^R$ and $C$ in the ansatz unless they are independent of $y$.  We make these statements more precise in section \ref{refAn}.  

Next we will use the equation to refine the metric ansatz which will require introduction of new coordinates we call canonical coordinates.  The reader can jump to section \ref{cansys} for the end product to avoid details.

\subsection{Refinining Ansatz}\label{refAn}

Now we use the equations in (\ref{om4}-\ref{jpsi}) to refine the metric ansatz in (\ref{ansatz}).  

We fix our conventions for the Hodge star operators.  The hodge star, $*_4$, on a form with $p$ legs on $\hat{x}$ plane and $q$ legs on $\tau$ plane can be written as
\begin{equation}
\hat{*}_4 X_p \wedge T_q =(-1)^p * X_p \wedge *_\tau T_q
\end{equation} where $*$ and $*_\tau$ act on $\hat{x}$ and $\tau$ planes, respectively.  They are defined as
\begin{align}
* d\hat{x}_1 &= - d\hat{x}_2, \qquad \;\; * d\hat{x}_2 = d\hat{x}_1 \\
*_\tau \eta_\tau &= -e^{C}\eta_\phi, \quad *_\tau e^{C} \eta_\phi = \eta_\tau  \\
* 1 &= e^{2A} dR_2\\
*_\tau 1 &= e^{2B+C} \eta_\tau \wedge \eta_\phi.
\end{align}  The exterior derivative on $\hat{x}$ plane is $\hat{d}$.  It useful to decompose the exterior derivative $d_4$ as
\begin{align}\label{exter}
d_4 &=  d_2 + \eta_\tau \partial_\tau + d\phi \partial_\phi \\
d_2 &= \hat{d} - e^C V^R \partial\tau.
\end{align} 

\subsubsection*{The $\Omega$ equations}
 
The ansatz for $\Omega$ trivially solves equation (\ref{ompsi}).  We, then, start with equation (\ref{omy}), which yields three conditions.  The first is
\begin{equation}
\partial_y C =0.
\end{equation}  This condition implies that we can set $C=0$.  To see this, first write $e^{-C}= \partial_\tau W$, for some $W$ that is independent of $y$.  Then we observe that
\begin{equation}
e^{-C} \eta_\tau = d W -\hat{d} W +V^R.  
\end{equation}  to complete the transformation, we shift $V^R$ by $\hat{d} W$ and recover the original form of $\eta_\tau$ with $W$ replacing $\tau$.  We also need to shift $B$ by $-C$ in order to completely remove $C$ from the metric.  The $\hat{x}$ coordinates stay the same but the derivative $\hat{d}$ gets shifted by a $\partial_\tau$ term.  This is cancelled by the shift in $V^R$ in the exterior derivative $d_4$ in (\ref{exter}).  Therefore we fix $C=0$ from now on.   

The next condition obtained from (\ref{omy}) is
\begin{equation}
V^I= V_0 - *V^R, \qquad \mbox{with} \qquad \partial_y V_0 =0. \label{defVI}
\end{equation} The last condition obtained from (\ref{omy}) is
\begin{equation}
y\partial_y \log \left(e^{2(A+B)} \cos^2(\zeta) \right) = -3 \tan^2(\zeta).
\end{equation}  We solve this condition by introducing functions $\Sigma$ and $\Lambda$ defined such that
\begin{equation}
e^{2A} = \frac{1}{3} \Sigma e^\Lambda, \qquad  e^{2B} =  \frac{y^3}{\Sigma \cos^2(\zeta)}. \label{defsgL}
\end{equation} The equation becomes
\begin{equation}
\cos^2(\zeta) = - \frac{3}{y\partial_y \Lambda}.  \label{defcosL}
\end{equation}

Now we look at equation (\ref{om4}).  This equation yields two conditions:
\begin{align}
\partial_\tau V_0 &= 0 \\
\rho &= \alpha d\phi + \frac{1}{2} *d_2 \Lambda - \frac{1}{2} \partial_\tau \Lambda \eta_\phi.  
\end{align}  The one-form $V_0$ depends only on the $\hat{x}$ coordinates.  

\subsubsection*{The $J$ equations}

Now we consider the $J$ equations.  These will yield the equations of motion for the system.  Equation (\ref{jpsi}) implies that the metric functions are independent of $\psi$, therefore it corresponds to a $U(1)$ isometry as expected.  The first set of non-trivial conditions are from the K\"{a}hler condition, equation (\ref{j4}).  These are
\begin{align}
d_2 V^R &=0 \label{dvr} \\
d_2 e^{2B} &= e^{2B} \partial_\tau V^R \label{db} \\
\partial_\tau e^{2A} dR_2 &= e^{2B} d_2 V^I. \label{tdifEL}
\end{align} The first condition, (\ref{dvr}), implies
\begin{equation}
d_2^2 =0, \qquad \mbox{thus} \qquad V^R = d_2 \Gamma = \frac{\hat{d} \Gamma}{1+ \partial_\tau \Gamma}
\end{equation} for some scalar function $\Gamma$.  Plugging this result into equation (\ref{db}) yields
\begin{equation}
d_2 \left(\frac{e^{2B}}{1+\partial_\tau\Gamma } \right) = 0.
\end{equation} We can then write 
\begin{equation}
e^{2B} = \frac{1}{3} G (1+ \partial_\tau \Gamma), \qquad \mbox{with} \qquad d_2 G=0.  
\end{equation}  The relations in (\ref{defsgL}) and (\ref{defcosL}) imply
\begin{equation}
\Sigma = -\frac{y^4 \partial_y \Lambda}{G (1+\partial_\tau \Gamma)}.
\end{equation}  Equation (\ref{tdifEL}) becomes
\begin{equation}
\frac{\partial_\tau}{1+\partial_\tau \Gamma} \left(\Sigma e^\Lambda \right) dR_2 = G d_2 V^I.
\end{equation}

Finally we can expand equation (\ref{j4}).  We collect $\rho$ and $J$ as 
\begin{align}
\rho &= \alpha d\phi + \frac{1}{2} *d_2 \Lambda - \frac{1}{2} \partial_\tau \Lambda \eta_\phi \\
J &= \frac{1}{3} \Sigma e^\Lambda dR_2 + \frac{1}{3} G (1+ \partial_\tau \Gamma) \eta_\tau \wedge \eta_\phi.
\end{align} We find
\begin{align}
\frac{1}{y} \partial_y \left[G(1+\partial_\tau \Gamma)\right] &= \partial_\tau^2 \Lambda \label{difB}\\
G(1+\partial_\tau \Gamma)\frac{1}{y} \partial_y  V^R &= d_2 \partial_\tau \Lambda \label{difVR}\\
G(1+\partial_\tau \Gamma)\frac{1}{y} \partial_y  V^I &=  \partial_\tau \Lambda \partial_z V^I - \partial_\tau * d_2 \Lambda \label{difVI}\\
\frac{1}{y} \partial_y  \left(\Sigma e^\Lambda \right)dR_2 &= \partial_\tau \Lambda d_2 V^I - d_2*d_2 \Lambda. \label{difEL}
\end{align}  The $V^I$ equation in (\ref{difVI}) is implied by (\ref{defVI}), (\ref{difB}) and (\ref{difVR}).  Equation (\ref{difVR}) can be written as
\begin{equation}
d_2 \left(y^2 \partial_\tau \Lambda - y\partial_y \Gamma G \right)=0.
\end{equation} We can therefore write
\begin{equation}
y^2 \partial_\tau \Lambda = y \partial_y \Gamma G + G_2 \qquad \mbox{where} \qquad d_2 G_2 =0.
\end{equation} Equation (\ref{difB}) becomes
\begin{equation}
\frac{\partial_\tau}{1+ \partial_\tau \Gamma} G_2 = \left(y \partial_y - \frac{y\partial_y \Gamma}{1+ \partial_\tau \Gamma} \partial_\tau \right) G.
\end{equation} 

This completes the reduction of the supersymmetry equations on the ansatz above.  However the story can be cleaned up more.  This follows from the fact that the twisted derivative operator $d_2$ is nilpotent; it defines new coordinates on the Riemann surface.  We study this next.

\subsubsection*{Canonical Coordinates}

We can simplify the system above by making the coordinate transformation
\begin{equation}
\log(q) = \tau + \Gamma, \quad s = y, \quad x_i = \hat{x}_i.  
\end{equation} In these coordinates we find
\begin{equation}
\partial_\tau = \frac{q \partial_q}{1-q \partial_q \Gamma}, \;\;\;  1+ \partial_\tau \Gamma = \frac{1}{1-q \partial_q \Gamma}.
\end{equation}  The other derivatives become
\begin{equation}
\begin{split}
d_2 &= dx_i \wedge \partial_i\\
s\partial_s &= y \partial_y - \frac{y\partial_y \Gamma}{1+ \partial_\tau \Gamma} \partial_\tau.
\end{split} \label{bcoord}
\end{equation} 

\subsection{Canonical System}\label{cansys}

The most general supersymmetric $AdS_5$ metric in M-theory that contains two circles fibred over a two-dimensional Riemann surface is
\begin{equation}\begin{split}
ds^2_{11} &= H^{-1/3}\left[ds^2_{AdS_5} + \frac{1}{3}  \frac{s^3(1-q\partial_q \Gamma)}{\Sigma G}\left(d\psi+\rho \right)^2 + \frac{1}{3} H ds^2_5 \right] \\
ds^2_5 &= \Sigma e^\Lambda \left(dx_1^2 + dx_2^2 \right) + \frac{G}{1-q\partial_q \Gamma} \left[   \frac{\Sigma}{s}\frac{ds^2}{s^2} + \eta_\tau^2+ \left(d\phi+V^I \right)^2 \right].
\end{split}
\end{equation}  We have set the AdS radius, $L$, to one.  We can reintroduce it by multiplying the metric by an overall $L^{4/3}$.  The forms are
\begin{align}
V^I &= V_0 - *d_2 \Gamma \\
\eta_\tau &= (1-q\partial_q\Gamma) \frac{dq}{q} - s\partial_s \Gamma \frac{ds}{s} \\
\rho &=\alpha  d\phi + \frac{1}{2} d_2 \Lambda- \frac{1}{2} \frac{q \partial_q\Lambda}{1-q\partial_q \Gamma}  \left(d\phi + V^I \right).
\end{align} The one-form, $V_0$, depends only on the Riemann surface coordinates, $x_i$.  The exterior derivative, $d_2$,  is taken along the $x_i$ directions.  The Hodge star operator acts as $*dx_1= -dx_2$.  
The metric functions are

\begin{align} 
H &= \frac{1}{4 s^2} \left[1- \frac{3s^3 \left(1-q\partial_q \Gamma \right)}{G\Sigma} \right] \\
\Sigma &= -\frac{s^3}G \left[(1-q\partial_q \Gamma) s\partial_s \Lambda + s\partial_s \Gamma q\partial_q \Lambda \right]. \label{sigdef}
\end{align}

The left over equations to solve are
\begin{align}
d_2 G_2 =d_2G &=0, \qquad s\partial_s G = q\partial_q G_2 \label{geqs}\\
s^2 q\partial_q \Lambda &= (1-q\partial_q \Gamma) G_2 + s\partial_s \Gamma G \label{qleq} \\
s\partial_s \left(\Sigma e^\Lambda\right) dR_2 &= G_2 d_2 V^I - s^2d_2 * d_2 \Lambda  \label{ssigeq}\\
q\partial_q \left(\Sigma e^\Lambda\right) dR_2 &= G d_2 V^I \label{qsigeq}
\end{align} where $dR_2= dx_1 \wedge dx_2$.

The $G$ equations can be solved in terms of a single function $X(s,q)$.  The solution is
\begin{equation}
G= q\partial_q X, \qquad G_2 = s\partial_s X. \label{defX}
\end{equation}

The system seems to be governed by three functions and a one-form.  Two of the functions, $\Lambda$ and $\Gamma$, depend on the Riemann coordinates and the interval coordinates $(s,q)$.  The third function $X$ depends only on the interval coordinates $(s,q)$.  The one-form, $V_0$, depends on the Riemann surface coordinates only.  The one-form can be set to zero if we allow for generic $\Gamma$.  One can do this by shifting $\Gamma$ by a $x$-dependent function and tune it such that its Laplacian cancels the contribution of $dV_0$ in equations (\ref{ssigeq}) and (\ref{qsigeq}).  Equivalently, we can keep $V_0$ and let it parametrize the part of $\Gamma$ that only depends on $x$.  We adopt this second choice.  

The function $\Lambda$ determines the Riemann surface while $\Gamma$ fixes the connections of the $U(1)$ fibrations.  We will call them the structure function and the embedding function respectively.  As we will below, the function $X$ determines the metric along the $(s,q)$ directions.  It will be used to find convenient coordinates for this plane.  So the system really governed by the two functions $\Lambda$ and $\Gamma$.  

\section{Finding solutions}\label{solve}

The goal is to understand the solution space of the system of differential equations in (\ref{geqs}-\ref{qsigeq}).  Generically, this is an homogeneity four problem.  Moreover the equations are second order and non-linear; the left hand sides of equations (\ref{ssigeq}-\ref{qsigeq}) involve derivatives of $\Sigma$ which itself is a derivative of other quantities as given by (\ref{sigdef}).  The system can be written in terms of a single function that satisfies a Monge-Amp\`{e}re equation.  The fact that the GMSW system is governed by a Monge-Ampere equation was first demonstrated in \cite{Lunin:2008tf}. Writing a general solution for this system is a tall task and we do not hope to achieve it here.    

We are going to look for solutions by making assumptions about the embedding function, $\Gamma$.  Most of the complications come from the fact that $\Sigma$ mixes the embedding and structure functions in a non-trivial way.  We can hope to find solutions if we can simplify this expression.  If we assume that the $x$-dependence of $\Gamma$ is through some implicit dependence on $\Lambda$ then $\Sigma$ simplifies.  It becomes an operator, that only depends on $(s,q)$, acting on $\Lambda$.  We can then pick coordinates where this operator is a simple derivative.  We can solve the system when $\Gamma$ is linear in $\Lambda$!  We make the ansatz
\begin{equation}
\Gamma = a \Lambda - \mathcal{Z}(s,q) + \log(q).  \label{gamdef}
\end{equation}  We could also add a term that depend only on $x$; however we know from the discussion below equation (\ref{defX}) that adding such term is equivalent to keeping $V_0(x)$. 

Now we define coordinates $(t,k)$ such that
\begin{align}
t \partial_t &= q\partial_q \mathcal{Z} s\partial_s - s\partial_s \mathcal{Z} q\partial_q \label{tcoord}\\
Gk\partial_k &=  \frac{f}{s^2} \left(a G s\partial_s - (aG_2 + s^2) q\partial_q \right). \label{kcoord}
\end{align} In the $(t,k)$ coordinates, we find
\begin{equation}
\begin{split}
f &= - k\partial_k X, \qquad \;\; s^2= 2T(t) -2a X \\
\Sigma  &= -\frac{s^3}{G} t\partial_t \Lambda, \qquad G= \frac{f}{s^2} g(t)
\end{split} \label{fundef}
\end{equation} where $X$ is defined in (\ref{defX}) and $g(t) = t\partial_t T$.  The functions $T(t)$ comes from integrating $\partial_k s^2$.   These coordinates are such that $\mathcal{Z}$ is independent of $t$, it defines the $k$ coordinates.  We fix it as $\mathcal{Z}= -\log{k}$.  Since the coordinate transformation does not involve the Riemann surface directions, the function $X$, defined in (\ref{defX}), depends only on $(t,k)$. 

The system of equations (\ref{qleq}-\ref{qsigeq}) becomes
\begin{align}
g(t) k\partial_k \Lambda &= -t \partial_t X \label{kllam} \\
k\partial_k \left(\Sigma e^\Lambda \right) &= k\partial_k X e^{2A_0(x)} \label{klam}\\
t\partial_t \left(\Sigma e^{\Lambda} \right) &= g(t) \Delta \Lambda + t\partial_t X e^{2A_0(x)}. \label{tlam}
\end{align} We have reduced the two-forms as
\begin{align}
d_2 V_0 &= e^{2A_0(x)} dx^1\wedge dx^2 \label{a0def}\\
d_2 * d_2\Lambda &= - \Delta \Lambda dx^1\wedge dx^2
\end{align} where $\Delta = \partial_{x_1}^2+ \partial_{x_2}^2$.  The function $\Lambda$ depends on all the coordinates.  We have introduced $A_0(x)$ to encode the $V_0$ data, it only depends on $x$ because $V_0$ depends only on the Riemann surface coordinates.  

Now, we study the system in (\ref{kllam}-\ref{tlam}). Equation (\ref{klam}) can be integrated to 
\begin{equation}
X e^{2A_0} + \frac{s^3}{G} t\partial_t e^{\Lambda} = L(x,t)
\end{equation} for some function $L$.  On the other hand equation (\ref{kllam}) implies that $\Lambda$ is separable as
\begin{equation}
\Lambda = D(x,t) + \Lambda_1(t,k). \label{lamsep}
\end{equation} These two conditions imply
\begin{equation}
L (x,t)= X e^{2A_0} +\frac{s^3}{G} t\partial_t e^{\Lambda_1}e^{D} +\frac{s^3}{G} e^{\Lambda_1} t\partial_t e^{D} . \label{classes}
\end{equation}  The function $X$ cannot be independent of $k$, otherwise $f$, as given in (\ref{fundef}), would vanish and the coordinate transformation in (\ref{kcoord}) would be degenerate.  This implies that when $e^{2A_0}$ is non-vanishing in equation (\ref{classes}), $e^D$ must be separable in $x$ and $t$.  If we want more generic solutions where $e^D$ is not separable, we must have $e^{2A_0}=0$.  The solution space splits into two classes.  These can further split depending whether $a$ in (\ref{gamdef})  vanishes or not.  After going through all possible scenarios, we find the following classes of solutions.
\begin{itemize}
\item[Class Ia] In this class of solutions, we impose $e^{2A_0}=0$.  Equation (\ref{a0def}) implies that the one-form, $V_0$, is flat.  We can set it to zero without lost of generality.  The warp factor for the Riemann surface, $\Lambda$, separates between $x$ and $k$ as implied by (\ref{kllam}).  We also impose $a=0$, this is equivalent to making $\Gamma$ independent of $x$.  The defining conditions are
\begin{equation}
\Lambda = D(x,t) + \Lambda_1(t,k), \qquad V^I =0.
\end{equation}  
\item[Class Ib] The solutions in this class satisfy the same conditions as in class Ia solutions except $a$ is non-vanishing.  The defining conditions are
\begin{equation}
\Lambda = D(x,t) + \Lambda_1(t,k), \qquad V^I =-a*d_2 D.
\end{equation}  
\item[Class II]  Finally we can consider solutions where $e^{2A_0(x)}$ is non-vanishing.  This requires $\Lambda$ to be separable in $x$ and $(t,k)$.  For this class, we can set $a=0$ without lost of generality.  This follows from the fact that if $a$ is non-zero, then $\Gamma$ will pick up a term that depends on $x$ only.  Such a term is already encoded in $V_0$ as discussed below equation (\ref{defX}).  We write the functions as
\begin{equation}
e^{\Lambda} = e^{2A(x)} e^{\Lambda_1(t,k)}, \quad e^{2A_0} = \kappa_2 e^{2A(x)}, \quad V^I = \kappa_2 V, \quad dV = e^{2A} dR_2.
\end{equation}
\end{itemize}

\subsection{The metric}\label{solmetric}

Before we study the different classes of solutions, we write the metric in the $(t,k)$ coordinates.  It is given as
\begin{align}
ds^2_{11} &= H^{-1/3} \left[ds^2_{AdS_5} + \frac{1}{9}  \left(1-4s^2H\right) \left(d\psi+\rho\right)^2 + \frac{1}{3} H ds^2_5 \right] \\
ds^2_5 &=\Sigma e^{\Lambda} \left(dx_1^2 + dx_2^2 \right) + g(t) \Sigma_0 \frac{dt^2}{t^2}  +f \left[\frac{dk^2}{k^2} +\frac{3g(t)}{s^2\Sigma_0\left(1-4s^2H\right)}   \left(d\phi+V^I \right)^2 \right]. \nonumber
\end{align}  The metric functions are
\begin{align}
\Sigma &= \frac{s^5}{g(t)  f} \Sigma_0, \qquad \Sigma_0 = - t\partial_t \Lambda \\
H &= \frac{1}{4s^2} \left(1 - \frac{3}{4} \frac{\left(t\partial_t s^2\right)^2 + 4a^2 f  g(t) \Sigma_0}{s^2 g(t) \Sigma_0} \right).
\end{align} The one forms are
\begin{align}
V^I &= V_0 -a *d_2 \Lambda \\
\rho &= p d\phi + \frac{1}{2} *d_2 \Lambda - \frac{3}{4} \frac{ t\partial_t X t\partial_t s^2 - 2af  g(t) \Sigma_0}{s^2 g(t) \Sigma_0 \left(1-4s^2 H\right)} \left(d\phi+ V^I\right).
\end{align}

We notice, from the metric and the equations of motion, that the function $g(t)$ can be removed by a coordinate transformation.  We will keep it explicit and fix it when it is convenient.  The choice of $g(t)$ will also fix the coordinate $t$.   

Solving the supersymmetry equations determines the six-dimensional internal manifold normal to the AdS space.  This geometry is a $S^1$ bundle over a five-dimensional base.  The base geometry is a $S^1$ bundle over a Riemann surface and two interval directions, $(t,k)$.  The conformal factor of the Riemann surface always separates into an $(x,t)$ and $(t,k)$ parts as discussed around equation (\ref{lamsep}).  The $(t,k)$ dependence determines the size of the Riemann surface on the tile while the $(x,t)$ dependence determines a one-parameter family of Riemann surface metrics along the $t$-direction.  The connection one-form of the $\psi$-circle fibration, $\rho$, has two parts.  The first is simply the spin connection of the Riemann at fixed $t$, while the second mixes the $\psi$-circle with $\phi$-circle.  The twisting varies along the interval.  Once the base metric is determined, $\rho$ is fixed.  

The connection of the $\phi$-circle fibration, $V^I$, determines the different classes of solution.  The supersymmetry equations in (\ref{ssigeq}-\ref{qsigeq}) and (\ref{klam}-\ref{tlam}) relate $V^I$ to the spin connection of the Riemann surface; more precisely to its variations along the intervals.  In class Ia solutions,  the $\phi$-circle fibration is trivial.  The system of equations will split into $(x,t)$ sector, which determines the family of Riemann surface metrics, and a $(t,k)$ sector which determines the metric along the intervals, the radius of the Riemann surface, and the shape and size of the $T^2$.  The $(t,k)$ dependence of metric can be solved exactly.  In class Ib solutions, we set $V^I$ proportional to the spin connection of the Riemann surface.  For this case, the system of equations also splits in manner similar to class Ia solutions.  In both classes, the conformal factor will satisfy a warped generalization of the $SU(\infty)$ Toda equation.  

In class II solutions, we consider the case when $V^I$ is constant on the interval directions.  The effect of this is to make the spin connection of the Riemann surface constant along the $(t,k)$ directions, and therefore the family of Riemann surface metric along $t$ collapse to one one of the constant curvature surface, $S^2$, $H_2$ or $T^2$.  We can then fix the Riemann surface metric to be the constant curvature one.  In this case, the problem of solving for the $(t,k)$-dependence of the metric is more non-trivial.  We discuss how to find them.   

Now we study the system of equations for the different cases and discuss how to solve them.  

\section{Class I  solutions: warped $SU(\infty)$ Toda systems}\label{classI}

The structure function for class Ia and Ib solutions separates as
\begin{equation}
\Lambda = D(x,t) + \Lambda_1(t,k).
\end{equation}  This, again, follows from equation (\ref{kllam}).  When $D$ is not separable in $x$ and $t$, equation (\ref{classes}), derived from (\ref{klam}), requires
\begin{equation}
e^{2A_0} =0, \qquad \mbox{and} \qquad \partial_k\left(\frac{s^3}{G} t\partial_t e^{\Lambda_1} \right)=\partial_k \left( \frac{s^3}{G} e^{\Lambda_1} \right)= 0. \label{L1c1}
\end{equation}  The vanishing of $e^{2A_0}$ implies that $V_0$ is flat, we can set it to zero without lost of generality.  The latter two constraints in (\ref{L1c1}) imply 
\begin{equation}
e^{\Lambda_1} = h_0(t) h_1(k), \qquad \mbox{and} \qquad G = h_0(t) h_1(k) s^3.  \label{seplg}
\end{equation}  In writing $e^{\Lambda_1}$  we have used the fact that the separability condition of $\Lambda$ is defined up an overall function of $t$.  We fix it such that $e^{\Lambda_1}$ is proportional to $h_0(t)$.  

Equation (\ref{tlam}) becomes
\begin{equation}
  g(t) \Delta D + t \partial_t \left[\frac{1}{h_0(t)} t\partial_t \left(h_0(t) e^{D} \right)\right]=0. \label{todaish1}
\end{equation} We call this equation warped $SU(\infty)$ Toda equation.  It is a generalization of $SU(\infty)$ Toda equation obtained, here, by fixing $g(t)=t^2$ and $h_0(t) \propto t^{-1}$.  The warping refers to the presence of $h_0(t)$.  

Differential equations for $h_0(t)$, $h_1(k)$ and $X$ can be obtained from equations (\ref{kllam}) and (\ref{seplg}) after we plug in for $G$ and $f$ as given in (\ref{fundef}).  Solutions to these equations will require some separability in $s^2$ and $X$.  Without lost of generality we can write these functions in terms $T(t)$ and a $k$-dependent function $P(k)$:
\begin{align}
\Sigma_0 &= -t\partial_t \log(h_0) - t\partial_t D \\
s^2 &=2 (\alpha_0 + \alpha_1 T(t)) (a_0 +a P(k)) \label{seps2}\\
X &= -c_0 - c_1 T(t) - c_2 P(k) - 2c_3 T(t) P(k) \label{sepX}\\
f &= \left(c_2 +2c_3 T(t) \right) k\partial_k P(k). \label{sepf}
\end{align} The relation between $s^2$ and $X$ in (\ref{fundef}) imply
\begin{equation}
\begin{split}
\alpha_0 a_0 &= ac_0, \qquad a \left(\alpha_0 -c_2 \right) =0 \\
\alpha_1  a_0-1&= ac_1, \qquad a \left(\alpha_1 -2c_3 \right) =0.  
\end{split} \label{alphacon}
\end{equation} After separating all the equations, we find
\begin{equation}
h_0(t) =  g(t) \frac{c_2+2c_3T(t)}{\left(\alpha_0 + \alpha_1 T(t) \right)^{5/2}}   \label{h0D}
\end{equation} for the $t$-dependence.  The $k$-dependence yields
\begin{align}
 k\partial_k \log h_1(k) &= c_1 + 2c_3 P(k) \label{Ghp}\\
k\partial_k P(k) &= h_1(k) \left(2a_0 + 2a P(k) \right)^{5/2}.  \label{Gpk}
\end{align} We can solve for $h_1$ in terms of $P(k)$ and obtain
\begin{equation}
h_1(k) = \left(2a_0 +2a P(k) \right)^{-3/2} \left[c_4 \left(2a_0 +2a P(k) \right)^{3/2}+ w_0+w_1 P+w_2 P^2 \right] \label{c1h1}
\end{equation} with
\begin{align}
aw_2 =0, \quad 2a_0 w_1 - 3aw_0 = c_1, \quad 4a_0 w_2 -aw_1 =2c_3. \label{wcons}
\end{align}

To continue, we need to specialize to different cases of $a$.

\subsection{Class Ia}

class Ia solutions further satisfy $a=0$.  The constraints on the parameters are
\begin{equation}
a=\alpha_0=w_0=0, \quad 2a_0 =1, \quad \alpha_1 = 2, \quad w_1 =c_1, \quad w_2 =c_3.
\end{equation} The $\alpha$'s are fixed by equations in (\ref{alphacon}); $a_0$ can be fixed without lost of generality.  The $w_i$'s are fixed by (\ref{wcons}); $w_0$ can be fixed without lost of generality.  The $c_i$ parameters are not constrained.

In order to write the metric, we need to fix $g(t)$.  It is convenient to chose
\begin{equation}
g(t)= t^2, \qquad \mbox{thus} \qquad s^2 =2T= \kappa_0 t^2_0 + t^2 
\end{equation} where $\kappa_0 = -1,0,1$.  We also shift $c_2$ as $c= c_2 + c_3 \kappa_1$ and fix $c_3=\kappa_2$.  Finally, we write $h_1(k) = k^2 e^{2U(k)}$.  The metric is then

\begin{align}
ds^2_{11} &=H^{-1/3} \left[ds^2_{AdS_5} + \frac{1}{3}  \frac{t^2}{ (t^2+\kappa_0 t^2_0)\Sigma_0} \left(d\psi+\rho\right)^2 + \frac{1}{3} H ds^2_5 \right] \label{classIa} \\
ds^2_5 &=\Sigma_0 \left[ dt^2 +e^{D} \left(dx_1^2 + dx_2^2 \right) \right]+(c + \kappa_2 t^2) e^{2U(k)}\left(dk^2+k^2 d\phi^2 \right). \nonumber
\end{align} The metric functions are given as
\begin{align}
\Sigma_0 &= - t\partial_t\log h_0(t) - t\partial_t D(x,t) \\
H &= \frac{\left(t^2+ \kappa_0 t_0^2\right)\Sigma_0-3t^2}{4(t^2+\kappa_0t_0^2)^2 \Sigma_0} \\
\rho &= (p + 1) d\phi + \frac{1}{2} *d_2 D(x,t) +k\partial_kU(k) d\phi.
\end{align} We can fix the charge of the holomorphic two-form in (\ref{omedef}) as $p=-1$ in order to remove the exact term in $\rho$.  
Equation (\ref{h0D}) becomes
\begin{equation}
h_0 = t^2  \frac{c+\kappa_2 t^2}{\left(t^2 + \kappa_0 t_0^2\right)^{5/2}}
\end{equation}  
From equations (\ref{Ghp}) and (\ref{Gpk}), we find
\begin{equation}
\Delta_k U(k) =\kappa_2 e^{U(k)}.
\end{equation} where $\Delta_k$ is the Laplacian on the $(k,\phi)$ plane.  The conformal factor of the Riemann surface satisfies the warped $SU(\infty)$ Toda equation:
\begin{equation}
\Delta D+ \frac{1}{t} \partial_t \left[ \frac{1}{h_0(t)} t\partial_t \left( h_0(t) e^{D}\right) \right] =0. \label{class1toda}
\end{equation}  

The space of solutions seem to have three free parameters $(c,\kappa_2,t_0)$.  We are free to fix two of these parameters up to signs.  Without lost of generality, we only consider cases when $\kappa_2 = -1,0,1$ and $t_0=1$.  Given the different choices for $\kappa_1$, we find six subclasses of one-parameter family of solutions.  

The $(k,\phi)$ plane parametrizes a second Riemann surface with curvature $-\kappa_2$ since the conformal factor $e^{2U}$ satisfies the Liouville equation.  At constant $t$, the internal geometry is then a $S^1$ bundle over a product of two Riemann surfaces.  The Riemann surface parametrized by $x$ mixes with $t$ to form a three-manifold similar to the eleven-dimensional LLM $AdS_5$ system \cite{Lin:2004nb}.  Its conformal factor satisfies a warped generalization of the $SU(\infty)$ Toda equation (\ref{class1toda}).  In \cite{Bah:2013ta} we analyse the space of these solutions, discuss their regularity conditions and how they generalize known solutions.  Next we show how the solutions of GMSW \cite{Gauntlett:2004zh}, which includes the M-theory uplift of the $Y_{p,q}$ \cite{Gauntlett:2004yd} and $\mathcal{N}=1$ Maldacena and N\'{u}\~{n}ez geometry \cite{Maldacena:2000mw}, embed into this class. 

\subsubsection{GMSW solutions}

The solutions of GMSW \cite{Gauntlett:2004zh} were obtained by considering cases when the internal six-dimensional geometry is complex.  Such solutions are $S^1$ bundles over two Riemann surfaces sitting on an interval.  Solutions of this type should embed into the class Ia.  The interval corresponds to the $t$-direction.  In GMSW, the conformal factors of the two Riemann surface satisfy the Liouville equation on their respective planes.  If we are to find them in class Ia, we need to look for solutions where $D$ is separable in $x$ and $t$.  We write
\begin{equation}
e^D = e^{2A(x)} L(t). 
\end{equation}  Equation (\ref{class1toda}) implies
\begin{equation}
\Delta A = \kappa_1 e^{2A}, \qquad t\partial_t \left(h_0(t) L(t) \right) = -\left(b_1 +\kappa_1 t^2\right) h_0(t).  
\end{equation} We solve for $L(t)$ and find
\begin{align} 
L(t) &= \frac{1}{3}\frac{t^2+\kappa_0}{t^2(c+ \kappa_2 t^2)} \hat{L}(t) \\
\hat{L}(t) &=  (c-\kappa_2 \kappa_0)(b_1-\kappa_1 \kappa_0)  +3 (\kappa_2 b_1 +\kappa_1 c-2\kappa_0\kappa_1\kappa_2) (t^2+\kappa_0) \nonumber \\ &+3 b_2 (t^2+\kappa_0)^{\frac{3}{2}} -3\kappa_2 \kappa_1(t^2+\kappa_0)^2.
\end{align} where the $b$'s are integration constants.  It is straightforward to write the metric functions, we obtain
\begin{align}
\Sigma_0 &= 3t^2 \frac{(b_1 +\kappa_1 t^2)(c+\kappa_2 t^2)}{(t^2+\kappa_0)\hat{L}(t)} \\
H &= \frac{(c+\kappa_2 t^2)(b_1+\kappa_1 t^2) - \hat{L}(t)}{4(\kappa_0+t^2)(b_1+\kappa_1 t^2)(c+\kappa_2 t^2)} \\
\rho &= *d_2 A + k\partial_k U(k) d\phi.
\end{align} The metric becomes
\begin{align}
ds^2_{11} &=  H^{-1/3} \left[ds^2_{AdS_5} + \frac{1}{9}  \frac{\hat{L}(t)}{(b_1 +\kappa_1 t^2) (c+\kappa_2 t^2)} \left(d\psi+\rho\right)^2 + \frac{1}{3} H ds^2_5 \right] \\
ds^2_5 &=3t^2 \frac{(b_1 +\kappa_1 t^2)(c+\kappa_2 t^2)}{(t^2+\kappa_0)\hat{L}(t)} dt^2 +(b_1+\kappa_1 t^2) ds^2\left(\mathcal{C}_g^1\right)+(c + \kappa_2 t^2) ds^2\left(\mathcal{C}_g^2\right) \nonumber
\end{align}  where $\mathcal{C}_g^i$ are the two Riemann surface with curvature $-\kappa_i$.  The known solutions are obtained by choosing the parameters in the following way
\begin{itemize}
\item The GMSW solutions \cite{Gauntlett:2004zh} are obtained by fixing $\kappa_0=0$.
\item The eleven-dimensional uplift of the $Y_{p,q}$ metrics are contained within the GMSW solutions \cite{Gauntlett:2004zh}.  This solutions is obtained by fixing $\kappa_1=0$ and $\kappa_2 =-1$, i.e. the first Riemann surface is a torus while the second is a two-sphere. 
\item The $\mathcal{N}=1$ Maldacena and N\'{u}\~{n}ez solution \cite{Maldacena:2000mw} is obtained within the GMSW solutions with $b_2=0$ and by fixing $\kappa_1=1$, $\kappa_2=-1$, i.e. the Riemann surface is a higher genus surface while the second is a two-sphere.  We also need to impose $b_1 = 3c$.  Finally the apparent free parameter, $c$, can be fixed by rescaling the $t$ coordinate.  
\end{itemize}

\subsection{Class Ib}

In class Ib solutions, $a$ is non-vanishing. We start by reducing the number of parameters.  Without lost of generality, we can fix $a_0=0$ by shifting the function $P(k)$ in (\ref{seps2}-\ref{sepf}).  This will also require us to redefine some of the parameters.  The constraints in (\ref{alphacon}) and (\ref{wcons}) imply
\begin{equation}
\alpha_0 = c_2, \quad \alpha_1 =2c_3=-aw_1, \quad c_0=w_2=0, \quad 3a^2w_0 = -ac_1 = 1. \label{ciipar}
\end{equation} We observe from equations (\ref{Ghp}), (\ref{Gpk}) and (\ref{c1h1}) that all $k$-dependent functions appearing in the metric are functions of $P(k)$.  This suggests that we should use $P$ as the actual coordinate instead of solving equation (\ref{Gpk}).  This equation will instead allows us to write $dk$ in terms of $dP$.  It is actually more convenient to introduce the coordinate $u$ from which we have
\begin{align}
P &= \frac{1}{2a} u^2, \qquad \qquad \qquad \frac{dk}k = \frac{3a}{h(u)} \frac{du}{u}\\
h_1(u) &=  \frac{1}{3a^2 u^3} h(u), \qquad h(u) = 1-3c_3 u^2 - b u^3 
\end{align} where $b=-3 a^2 c_4$.  

For this class, we can fix $2T=g(t)=t^2$.  The metric functions are 
\begin{align}
\mathcal{T} &= 9c_3^2  u^2 t^2 + 3\left(c_2+c_3 t^2\right) h(u) \Sigma_0 \\
H 
&=\frac{1}{4\left(c_2+c_3 t^2 \right)} \left[bu + 3c_3 \frac{\left(c_2+c_3 t^2 \right) \Sigma_0-c_3  t^2}{\left(c_2+c_3  t^2\right) \Sigma_0} \right].
\end{align} The metric is given as
\begin{align}
ds^2_{11} &= H^{-1/3} \left[ds^2_{AdS_5} + \frac{1}{9} \frac{\mathcal{T}}{3\left(c_2+c_3 t^2 \right) \Sigma_0}\left(d\psi+\rho\right)^2 + \frac{1}{3} H ds^2_5 \right] \label{class2metric}\\
ds^2_5 &=\Sigma_0\left[dt^2+ e^{D} \left(dx_1^2 + dx_2^2 \right) \right]+3\left(c_2+c_3 t^2 \right) \left[\frac{du^2}{h(u)} + \frac{t^2h(u)}{a^2\mathcal{ T}}   \left(d\phi- a *d_2 D\right)^2 \right]. \nonumber
\end{align}  The warp factor of the Riemann surface satisfies the $SU(\infty)$ Toda equation:
\begin{equation}
\Delta D+ \frac{1}{t} \partial_t \left[\frac{1}{h_0}t\partial_t \left(h_0 e^{D} \right) \right] =0 \label{class2toda}
\end{equation} with
\begin{align}
h_0(t)&=  t^2 \left(c_2 +  c_3 t^2 \right)^{-3/2} \\
\Sigma_0 &= - t\partial_t \log h_0(t) - t\partial_t D \\
\rho &= \left(p+\frac{1}{2a} \right) d\phi - \frac{9}{2a} \frac{c_3 t^2}{\mathcal{T}}\left(d\phi-a *d_2 D \right).
\end{align}

It is clear from the metric that we can fix $a=1$ without lost of generality; this requires rescaling the $\phi$ coordinate.  We can also fix $c_3$ and $c_2$ up to overall signs, therefore we can consider cases where $3c_3 = -1,0,1$ and $c_2=-1,0,1$ with out lost of generality.  The only free parameter of the system is $b$.  

Next we show how the LLM solutions fit in this system\footnote{The embedding of LLM into the GMSW system (in section \ref{GMSWsys}) was also understood by Lunin in \cite{Lunin:2008tf}.}.  

\subsubsection{LLM solutions}

The LLM solutions are obtained by fixing $b=0$, $c_2=0$ and $3c_3=1$.  In this section we work out the metric explicitly.  Fixing $c_2=0$ implies that the conformal factor of the Riemann surface satisfies the $SU(\infty)$ Toda equation in (\ref{class2toda}) as expected.  The LLM solutions contain a topological $\widetilde{S}^4$ with $U(1)\times SU(2)$ isometry corresponding to the $\mathcal{N}=2$ R-symmetry.  The interval for this $\widetilde{S}^4$ is $u$.  The metric along the $(\phi,\psi)$ directions, in (\ref{class2metric}), should diagonalize to two circles corresponding to the $\mathcal{N}=2$ $U(1)$ R-symmetry and the Cartan of the $SU(2)$ R-symmetry.   The metric is diagonalized by 
\begin{equation}
\hat{\phi}= \frac{3}{2} \left(p+\frac{1}{2} \right) \phi + \frac{3}{2} \psi, \qquad \chi = \phi - \hat{\phi}.
\end{equation}  We can make this transformation even when $c_2\neq 0$ to obtain
\begin{align}
ds^2_{11} &=  H^{-1/3} \left[ds^2_{AdS_5} +\frac{t^2}{4(3c_2+t^2) \Sigma_0} \left(d\chi-*d_2 D\right)^2 + \frac{1}{3} H ds^2_5 \right] \\
ds^2_5 &=\Sigma_0\left[dt^2+ e^{D} \left(dx_1^2 + dx_2^2 \right) \right]+\left(3c_2+ t^2 \right) \left[\frac{du^2}{1-u^2} + (1-u^2)  d\hat{\phi}^2\right]. \nonumber
\end{align} This matches the LLM metric as described by Gaiotto and Maldacena \cite{Gaiotto:2009gz} for $c_2=0$.  

\section{Class II solutions: Liouville systems}\label{classII}

In class II solutions, the conformal factor of the Riemann surface is separable between $x$ and other coordinates.  The $x$ dependent part satisfies the Liouville equation.  For this class we can fix $a=0$ without lost of generality.  We can write
\begin{equation}
e^{\Lambda} = e^{2A(x)}e^{\Lambda_1(t,k)}, \qquad \Delta A = \kappa_1 e^{2A}, \qquad e^{2A_0(x)} = \kappa_2 e^{2A(x)}  \label{classii}
\end{equation}  where $\kappa_i$ are constants.  The curvature of the Riemann surface is $-\kappa_1$.  In writing the differential equations, it is more useful to see things as functions of $T(t)$.  When we fix $g(t)$, we would have also fixed $T(t)$ and therefore the $t$-dependence of the system.  We switch $t$ derivatives to $T$ derivatives as $t\partial_t = g(t) \partial_T$.  
The supersymmetry equations in (\ref{kllam}-\ref{tlam}) become
\begin{align}
\partial_T e^{\Lambda_1} &=\frac{1}{\left(2T\right)^{5/2}} k\partial_k X \left( \kappa_2 X +L_0(t) \right) \label{L1c2}\\
 k\partial_k e^{\Lambda_1}&= -\partial_T X e^{\Lambda_1} \label{l1c2}
\end{align} Equation (\ref{tlam}) imply
\begin{equation}
L_0(t) = \ell_0 +2 \kappa_1 T
\end{equation} where $\ell_0$ is a constant.

The goal is to write metric solutions.  This problem does not require us to explicitly solve the equations in (\ref{L1c2}) and (\ref{l1c2}).  We need to write a metric that is consistent with the equations.  We saw a little bit of this when we worked out class Ib solutions.  There, we obtained equations (\ref{Ghp}-\ref{c1h1}) for $P(k)$ and $h_1(k)$.  We observed, as discussed below equation (\ref{ciipar}), that we should use $P(k)$ as the coordinate instead of $k$ since all metric functions depended on $k$ through $P(k)$.  The differential equation was then used as the Jacobian of transformation from $k$ to $P$ in the metric.  We use this trick at industrial scale to write solutions for class II.  We present an algorithm for doing this and work out an example that includes B$^3$W solutions \cite{Bah:2012dg}.  In \cite{Bah:2013ta} we do a more extensive study of class II solutions.  

Start by introducing a third coordinate $u$ that depends on $(t,k)$.  We assume that both $X$ and $e^{\Lambda_1}$ are polynomials in $u$ with $t$-dependent coefficients.  We denote them as
\begin{equation}
X = \sum X_n(t) u^n, \qquad e^{\Lambda_1} =\frac{1}{\left(2T\right)^{3/2}} \sum P_n(t) u^n. \label{XLans}
\end{equation} It is convenient to factor out an overall $(2T)^{-3/2}$ in $e^{\Lambda_1}$ in order to cancel the $(2T)^{-5/2}$ factor in equation (\ref{L1c2}).  The explicit form of $u$ is not important, however when we expand the equations above, $k$ and $t$ derivatives of $u$ will appear by the chain rule.  We also assume that these functions are polynomials in $u$ with $t$-dependent coefficients.  We denote them as
\begin{equation}
\begin{split}
k\partial_k u &= -D_k(u,t) =-\sum C_n(t) u^n \\
\partial_T u &= D_t(u,t)= \sum T_n(t) u^n. \end{split} \label{intD}
\end{equation}  The integrability condition for $u$ implies
\begin{equation}
\left. \frac{\partial}{\partial T}\right|_u D_k = D_k \partial_u D_t - D_t \partial_u D_k.
\end{equation} The $T$-derivative on the left is taken at fixed $u$.  This relation constrains possible choice for the $C_n$'s once given the $T_n$'s.  

The next step is to plug the ansatz in (\ref{XLans}) into equations (\ref{L1c2}) and (\ref{l1c2}).  We expand these equations in powers of $u$ by using (\ref{intD}).  This yields differential equations for the $X_n$'s and $P_n$'s in terms of the $C_n$'s and $T_n$'s.  When this system is solvable, we can write a metric in $(u,t)$ coordinates by replacing $dk$ with
\begin{equation}
\frac{dk}{k} = -\frac{du}{D_k(u,t)} + \frac{g(t)D_t(u,t)}{D_k(u,t)} \frac{dt}{t}.
\end{equation} The metric can be written as
\begin{align}
ds^2_{11} &= H^{-1/3} \left[ds^2_{AdS_5} + \frac{1}{9} \frac{(2T)^{3/2} e^{\Lambda_1}}{D_k \partial_u X \left(L_0+\kappa_2 X \right)} \left(d\psi+\rho\right)^2 + \frac{1}{3} H ds^2_5 \right] \label{classIImetric} \\
ds^2_5 &=\left(L_0+\kappa_2 X\right) \left[e^{2A}\left(dx_1^2 + dx_2^2 \right) + \frac{D_k \partial_u X }{(2T)^{3/2} e^{\Lambda_1}} \frac{ g^2(t) dt^2}{2t^2 T} \right]\nonumber \\
&+ D_k \partial_u X \left(d\phi-\kappa_2 *d_2 A_0\right)^2 + \frac{\partial_u X}{D_k} \left(du - D_t \frac{g(t) dt}{t} \right)^2.  \nonumber
\end{align} The functions are
\begin{align}
H &= \frac{1}{8T} \left[\frac{D_k \partial_u X \left(L_0 + \kappa_2 X \right) - 3(2T)^{3/2} e^{\Lambda_1}}{D_k \partial_u X \left(L_0 + \kappa_2 X \right)} \right] \\
\rho &= \alpha d\phi + *d_2 A - \frac{1}{2} D_k \partial_u \Lambda_1 \left(d\phi-\kappa_2 *d_2 A \right).
\end{align}

The solutions found in this way tend to have many parameters coming from integration constants.  Moreover at various steps of reducing the equations, the system breaks into subclasses.  In order to illustrate these points, and the validity of this method, we consider an example that leads to B$^3$W solutions \cite{Bah:2012dg}.  

\subsection{Example}

First we chose the $X_n$'s and $T_n$'s as
\begin{align}
T_n(t) &= \frac{b}{X_1(t)} \alpha_{n+1} \\
X_0 &= c_0 +c_1 T \\
X_1(t) &= c_2 +2c_3 T.
\end{align} The $\alpha$'s are non-vanishing only for $n={-1,0,1}$; $b$ is also constant.  When we plug this choice for $T_n$ into (\ref{intD}), we obtain several solutions for the $C_n$'s.  We restrict to one of simplest where they are independent of $t$.  We have
\begin{equation}
D_k(u,t) = \frac{1}{u} K(u) = \frac{1}{u} \left(\alpha_0 + \alpha_1 u + \alpha_2 u^2 \right).
\end{equation}

Equation (\ref{l1c2}) implies
\begin{equation}
K(u)  \partial_u \Lambda_1 = c_1 u +2c_3 u^2+ bK(u).  \label{KuL}
\end{equation}  The function $e^{\Lambda_1}$ must be separable in $u$ and $t$.  We write it as
\begin{align}
e^{\Lambda_1} &= \frac{1}{(2T)^{3/2}} h_0(t) h_1(u) \\
K(u)  \partial_u \log h_1(u) &= c_1 u +2c_3 u^2+ bK(u). \label{kh1u}
\end{align} Now we expand equation (\ref{L1c2}) to obtain
\begin{align}
\frac{K(u)}{h_1(u)} X_1^2\left(L_0 +\kappa_2 X_0 + u\kappa_2 X_1 \right) &=  -2b(2c_3 + b\alpha_2) h_0 T u^2 - 2\alpha_0 b^2 T h_0  \label{tLex} \\
&- u \left[X_1 (-3h_0 + 2T\partial_T h_0) +2(c_1 +b\alpha_1)b h_0 T \right]. \nonumber
\end{align}

Solutions to (\ref{tLex}) require $h_1(u)$ to be a ratio of two polynomials.  This is compatible with (\ref{kh1u}) only when 
\begin{equation}
2c_3 +\alpha_2 b=0 
\end{equation}  in order to match the highest power of $u$ on the left hand side.  This constraint removes the $u^2$ term on the right hand side of (\ref{tLex}), therefore the generic solution of $h_1(u)$ is a ratio of two polynomials with degrees $m+2$ and $m$, respectively.  We consider the simple case where $h_1(u)$ is quadratic in $u$ and write it as
\begin{equation}
h_1(u) = \beta_0 + \beta_1 u + \beta_2 u^2.
\end{equation}  Equation (\ref{kh1u}) restricts the parameters as
\begin{equation}
\begin{split}
 2\beta_0\left(\beta_2\alpha_0 -\alpha_2 \beta_0\right) &= \beta_1 \left(\alpha_0 \beta_1-\beta_0 \alpha_1 \right), \qquad \beta_1 = b \beta_0 \\
  2\beta_0\left(\beta_2\alpha_1 -\alpha_2 \beta_1\right) &= \beta_1 \left(\alpha_0 \beta_2-\beta_0 \alpha_2 \right), \qquad c_1 = 2\alpha_2 -b\alpha_1.
\end{split} \label{kh1cons}
\end{equation}

 The $u^0$ and $u^3$ terms of equation (\ref{tLex}) yield
\begin{align}
h_0(t) &= -\frac{1}{2\beta_0 b^2}X_1^2 \frac{L_0+\kappa_2 X_0}{T} \label{h0u0}\\
2 T \partial_T h_0   &= -\frac{1}{2\beta_2 \beta_0 b^2 T} X_1^2\left(2\beta_0 \alpha_2 b^2\kappa_2 T + 3\beta_2(L_0+\kappa_2X_0) \right) \nonumber \\ & + \frac{(c_1+b\alpha_1)}{\beta_0 b} X_1 (L_0 +\kappa_2 X_0). \label{h0u3}
\end{align} The $u^1$ and $u^2$ terms yield
\begin{align}
\beta_2\left(\alpha_1 \beta_0 -\alpha_0 \beta_1  \right) \left(L_0 + \kappa_2 X_0 \right) &= \beta_0 \left(\alpha_2\beta_0 -\alpha_0 \beta_2 \right) \kappa_2 X_1 \label{h0u1}\\
\beta_2\left(\alpha_2 \beta_0 -\alpha_0 \beta_2  \right) \left(L_0 + \kappa_2 X_0 \right) &= \beta_0 \left(\alpha_2 \beta_1-\alpha_1\beta_2 \right)\kappa_2 X_1. \label{h0u2}
\end{align}  

Plugging (\ref{h0u0}) into (\ref{h0u3}) implies
\begin{equation}
\ell_0 + \kappa_2 c_0 =0, \qquad \mbox{and} \qquad 3\beta_2 \left(2\kappa_1+ \kappa_2 c_1\right) = -2\beta_0 \alpha_2 \kappa_2 b^2
\end{equation}  and therefore
\begin{align}
h_0(t) &= \frac{\alpha_2}{3\beta_2} \kappa_2 X_1^2 \\
L_0+ \kappa_2 X_0 &=  -\frac{2\beta_0 \alpha_2 \kappa_2 b^2}{3\beta_2} T.
\end{align}

One can check that when $\alpha_i\beta_j \neq \beta_i\alpha_j$, the equations in (\ref{h0u1}) and (\ref{h0u2}) are not compatible with the constraints in (\ref{kh1cons}) unless $\beta_1=\beta_0=0$.  We must have
\begin{equation}
\alpha_i\beta_j = \beta_i\alpha_j \qquad \mbox{or} \qquad \beta_1=\beta_0 =0.
\end{equation}  The solution space splits into two types.  We can write the general metric for this example as
\begin{align}
ds^2_{11} &= H^{-1/3} \left[ds^2_{AdS_5} + \frac{1}{3} \frac{h_1(u)}{K(u)} \frac{\alpha_2 uX_1}{\mathcal{T}} \left(d\psi+\rho\right)^2 + \frac{1}{3} H ds^2_5 \right] \\
ds^2_5 &=\frac{\kappa_1}{3\beta_2} \mathcal{T} e^{2A} \left(dx_1^2 + dx_2^2 \right) +\frac{1}{u} K(u) X_1 \left(d\phi-\kappa_2 *d_2 A \right)^2 \nonumber \\
&  + \frac{K(u)}{h_1(u)} \frac{\mathcal{T}}{\alpha_2 uX_1} \frac{g^2(t)dt^2}{2t^2 T} + \frac{uX_1}{K(u)} \left(du -\frac{bK(u)}{u X_1} \frac{g(t) dt}{t} \right)^2. \nonumber
\end{align}  The metric functions are
\begin{align}
X_1 &= c_2 -2\alpha_2 b T \\
\mathcal{T} &= 3\beta_2 u X_1- 4 \beta_0 \alpha_2 b^2 T\\
H &= \frac{1}{8T} \frac{3uX_1 \left[\beta_2 K(u) - \alpha_2 h_1(u) \right] - 4\beta_0b^2  \alpha_2 K(u)T}{K(u) \mathcal{T}} \\
\rho &= \alpha d\phi + *d_2 A - \frac{1}{2u} \left(\alpha_0 b + 2\alpha_2 u\right) \left(d\phi - \kappa_2 * d_2 A \right).
\end{align} 

Now we reduce to B$^3$W solutions.

\subsubsection{B$^3$W Solutions}

We restrict to
\begin{equation}
\alpha_i \beta_j = \beta_i \alpha_j, \qquad \mbox{thus} \qquad  \beta_2 K(u) = \alpha_2 h_1(u).
\end{equation}  The metric becomes
\begin{align}
ds^2_{11}&= H^{-1/3} \left[ds^2_{AdS_5} + \frac{\kappa_2 B}{12} e^{2A} \left(dx_1^2 + dx_2^2 \right)  + \frac{B}{4} \frac{g^2(t)dt^2}{2ut^2X_1T} + \frac{1}{3} uX_1 Hds^2_3 \right] \\
ds^2_3 &= \frac{1}{K(u)} \left(du-\frac{\alpha_1K(u)}{\alpha_0uX_1} \frac{g(t) dt}{t} \right)^2+ \frac{K(u)}{u^2} \left(d\phi - \kappa_2 *d_2A\right)^2+ \frac{4}{3B} \left(d\psi+\rho \right)^2\nonumber
\end{align} where 
\begin{equation}
B = -\frac{\alpha_1^2}{3\alpha_0} \qquad X_1 = c_2 -\frac{\alpha_2 \alpha_1}{\alpha_0} T, \qquad H = \frac{B}{8B T + 4u X_1}.
\end{equation} The Riemann surface data is encoded in $A(x)$ which satisfies
\begin{equation}
\Delta A = \kappa_1 e^{2A}, \qquad \mbox{where} \qquad 2\kappa_1 = -\kappa_2 \left(B  + 2\alpha_2  \right).
\end{equation}The one-form $\rho$ is given as
\begin{equation}
\rho = \alpha d\phi + *d_2 A - \frac{1}{2u} \left(\alpha_1 + 2\alpha_2 u\right) \left(d\phi - \kappa_2 * d_2 A \right).
\end{equation}  The solution of B$^3$W corresponds to fixing $\alpha_0 = -\frac{1}{36}$, $\alpha_1 =1$ and $c_2=3$.  We also fix $g(t)=t^2$ and $2T=t^2$.  This matches the solution as described in Appendix D of \cite{Bah:2012dg}.  

\section{Summary and Discussion}\label{discussions}

Our goal in this paper was to understand supersymmetric $AdS_5$ solutions in M-theory when the internal space contains a two-dimensional Riemann surface, $\mathcal{C}_g$, and admits at least an additional $U(1)^2$ isometry.  The six-dimensional internal geometry is generically a $T^2$ bundle over $\mathcal{C}_g$ with two intervals that form a two-dimensional subspace.  The size and shape of the $T^2$ can vary on the interval directions.  The system is governed by two functions $\Lambda$ and $\Gamma$ that dependent on the Riemann surface coordinates $(x_1,x_2)$ and interval directions.  The metric on $\mathcal{C}_g$ is conformal to $\mathbb{R}^2$ with $e^\Lambda$ as conformal factor.  The circle coordinates on the $T^2$ are $\psi$ and $\phi$. The connection for the $\phi$-circle fibration is $V^I=-*d_2\Gamma$ (star and derivative are taken on the Riemann surface).  The connection for the $\psi$-circle fibration is the spin connection of $\mathcal{C}_g$ plus a $\phi$ mixed term corresponding to the off diagonal term of the metric on the $T^2$.   The supersymmetry conditions reduce to a system of second order non-linear equations for $\Lambda$ and $\Gamma$ in all four coordinates.  It is solvable when we make certain identifications with $\Gamma$.  We organize them into three classes.  The coordinates on the interval directions are $t$ and $k$.  

\subsubsection*{class Ia solutions}

 For class Ia, we assume that $\Gamma$ is constant on $\mathcal{C}_g$.  The effect of this is to trivialize the $\phi$-circle fibration, i.e. $V^I=0$.  The equations reduce such that the $\phi$-circle joins with $k$ to form a second Riemann surface with constant curvature, $\mathcal{C}'_{g'}$.  The original Riemann surface, $\mathcal{C}_g$, joins with the $t$-interval to form a three-manifold that describes a one-parameter family of Riemann surface metrics.  The conformal factor separates into a function that depends on $(t,k)$, which goes into fixing the size of $\mathcal{C}_g$ along the interval directions, and another that depends on $(x,t)$.  The full metric is determined up to the $(x,t)$-dependent part of the conformal factor, denoted as $D(x,t)$.  The $[t]\times \mathcal{C}_g$ part of the metric is 
\begin{equation}
ds^2_6 = ... -t\partial_t \log\left(h_0(t) e^D\right)\left[dt^2 +e^D \left(dx_1^2 + dx_2^2 \right) \right]+ ... \label{submet}
\end{equation} The ellipsis correspond to overall warping and the other parts of the metric.  The function $D(x,t)$ satisfies a warped generalization of the $SU(\infty)$ Toda equation:
\begin{equation}
\Delta D + \frac{1}{t} \partial_t \left[ \frac{1}{h_0(t)} t\partial_t \left(h_0(t) e^D\right) \right]=0 \label{todagen}
\end{equation} where $\Delta= \partial_{x_1}^2+\partial_{x_2}^2$.  The function $h_0(t)$ is known.  The total internal space is a $S^1$ bundle over $[t]\times\mathcal{C}_g \times\mathcal{C}'_{g'}$.  The connection of the $\psi$-circle fibration is completely fixed by the spin connection of $\mathcal{C}_g \times \mathcal{C}_g'$.  The degree of the fibration only depends on the period of $\psi$ which is fixed by regularity conditions.  

In the special case when $e^{D}$ is separable, we can write $D = 2A(x) + D_0(t)$.  Equation (\ref{todagen}) implies that the $x$ dependent part of $D$ satisfies the Liouville equation
\begin{equation}
\Delta A(x) = \kappa_1 e^{2A(x)}
\end{equation} where $\kappa_1 = -1,0,1$.  The solutions correspond to the constant curvature Riemann surfaces $\mathbb{H}_2$ ($\kappa_1=1$), $T^2$ ($\kappa_1=0$) and $S^2$ ($\kappa_1=-1$).  The $\mathbb{H}_2$ can be made compact by mod-ing with a Fuchsian subgroup, of the $SL(2,\mathbb{R})$ isometry, to obtain a genius $g>1$ Riemann surface.  The $t$-dependent part contributes to the size of $\mathcal{C}_g$ along the intervals.  This subclass includes the solutions of GMSW \cite{Gauntlett:2004zh} and therefore the eleven-dimensional uplift of the $Y_{p,q}$ solutions \cite{Gauntlett:2004zh},\cite{Gauntlett:2004yd} and $\mathcal{N}=1$ Maldacena and N\'{u}\~{n}ez solution \cite{Maldacena:2000mw}.  

\subsubsection*{class Ib Solutions}

class Ib solutions are obtained by identifying the $x$-dependence of $\Gamma$ with $\Lambda$.  This fixes the connection for the $\phi$-circle fibration as $V^I = -*d_2\Lambda$.  The supersymmetry equations still imply that $e^\Lambda$ is separable as functions of $(t,k)$ and $D(x,t)$.  The part of the metric that include the Riemann surface has the same form as (\ref{submet}) and (\ref{todagen}) with a different $h_0(t)$.  Generically the solution is a $T^2$ bundle over $\mathcal{C}_g \times [t] \times [k]$.  The connections for the circle-fibrations are completely fixed by supersymmetry in terms of the spin connection of $\mathcal{C}_g$.  

In the special case when $h_0(t) \propto t^{-1}$, equation (\ref{todagen}) reduces to the $SU(\infty)$ Toda equation.  In one of the sub-sectors of the solutions we can find circle coordinates that diagonalize the metric on $T^2$.   One of the circle stays non-trivially fibred on $\mathcal{C}_g$ while the other joins with $k$ interval to form a two-sphere.  This solution is precisely the eleven-dimensional $AdS_5$ solution of LLM \cite{Lin:2004nb}.  When $e^D$ is separable, the Riemann surface reduces to the constant curvature ones.  The regular solution, which picks out the negatively curved Riemann surface \cite{Gaiotto:2009gz}, corresponds to the $\mathcal{N}=2$ Maldacena and N\'{u}\~{n}ez solution \cite{Maldacena:2000mw}.

\subsubsection*{Class II Solutions}

Class II solutions are obtained when $e^\Gamma$ is a separable in $x$ and $(t,k)$.  With this choice, the supersymmetry equations forces $\Lambda$ to also separate into a sum of a $x$-dependent function and $(t,k)$-dependent function. The $x$-dependent part will satisfy the Liouville equation and therefore $\mathcal{C}_g$ can be taken as one of the constant curvature Riemann surfaces.  This choice for $\Gamma$ also fixes the connection for the $\phi$-circle fibration to $V^I = -\kappa_2 V$ where $V$ is the spin connection on $\mathcal{C}_g$.  The parameter $\kappa_2$ labels different solutions, it determines the ratio of the degree of the two circle-fibrations.  

All equations in class II solutions reduce to a system on the $(t,k)$ directions that is far less constrained than in class Ia and Ib.  We can find many different solutions including the separable ones of class Ia and Ib.  For this case, we presented a general algorithm for writing metrics.  The general form of the solution is a $T^2$ bundle over $\mathcal{C}_g \times [t] \times [k]$.  We work out an example and show that it includes the solutions of B$^3$W \cite{Bah:2012dg}.  In \cite{Bah:2013ta}, we do a more careful study of these solutions and discuss their regularity conditions.  

\subsubsection*{Punctures}

The systems and solutions discussed above correspond to the near-horizon geometry of a stack of M5-branes wrapping $\mathcal{C}_g$ inside a larger structure of intersecting branes.  The field theory dual describe such structure from the point of view of the M5-branes on $\mathcal{C}_g$.  This interpretation of $AdS_5$ solutions in M-theory was made precise in the case of LLM solutions \cite{Lin:2004nb} by Gaiotto and Maldacena (GM) \cite{Gaiotto:2009gz} as reviewed in the introduction.  Our goal in upcoming works is to make this interpretation precise for the solutions discussed above.  The strategy is to start with a seed solution like the MN2 solution.  The Riemann surface in the seed solution must have trivial relative warping with respect to the $AdS_5$ geometry.  Next, we interpret the rest of the solutions as emerging from adding localized sources on the Riemann surface.  In class Ia solutions, the seed solution is the $\mathcal{N}=1$ Maldacena and N\'{u}\~{n}ez solution \cite{Maldacena:2000mw}.  In class Ib solutions, the seed is  the MN2 solution.  In class II solutions there is a one parameter family of seed solutions which include the MN solutions, these are the B$^3$W solutions \cite{Bah:2012dg}.   

It is interesting to notice that class Ib solutions contain more than just LLM.  Since the seed is always MN2, if these non-LLM solutions exist, it is reasonable to expect punctures that break the $\mathcal{N}=2$ supersymmetry down to $\mathcal{N}=1$.  These would be $\mathcal{N}=1$ punctures of $\mathcal{N}=2$ class $\mathcal{S}$ SCFT's.  It would be interesting to understand how these defects work from the point of view Gaiotto's classification.  It is known that the CFT dual to MN1 solution is the IR limit of the mass deformed field theory dual to MN2 \cite{Benini:2009mz}.  It is natural to wonder whether class Ia solutions describe the gravity dual to mass deformations of Gaiotto theories in general.  It is tempting to expect this given the similarity between class Ia, and class Ib solutions.  This will require understanding the solutions of the warped $SU(\infty)$ Toda equations.  

Finally we observe that the separable sector of class Ia and class Ib solutions, and all of the class II solutions have the constant curvature Riemann surface.  Except for the seed solutions, there is always a relative warping between $\mathcal{C}_g$ and $AdS_5$.  We expect these solutions to emerge from punctures.  Since we have the constant curvature Riemann surface, these source must be uniformly distributed and their density function should be related to this relative warping.  In other words, we have smeared punctures.  We explore these objects in \cite{Bah:2013ta}.  

\subsubsection*{To IIA and IIB}

In all three classes there exist solutions with shrinking $T^2$, which is unrelated to $T^2$ from the bundle.  One such example is the $Y_{p,q}$ in class Ia.  In fact the $Y_{p,q}$ were discovered in M-theory by GMSW and then studied them in IIB.  In class II there are more examples when we fix the Riemann surface to be $T^2$.  One should compactify down to IIA supergravity and then T-dualize to IIB.  This should yield IIB metric including the $L(p,q,r)$ solutions \cite{Cvetic:2005ft}.  It is interesting to wonder whether there are more examples of Sasaki-Einstein metrics than the $L(p,q,r)$ solutions in IIB.


\acknowledgments

I am grateful to Phil Szepietowski and Maxime Gabella for reviewing earlier draft.  I would also like to thank Brian Wecht, Nikolay Bobev, Chris Beem, Leopoldo Pando Zayas, James Liu, Iosif Bena, Nicholas Warner, Daniel Waldram, Mariana Gra\~{n}a and Nicholas Halmagyi for useful discussions.  IB is supported in part by ANR grant 08-JCJC-0001- 0 and the ERC Starting Grants 240210 - String-QCD-BH and 259133 - ObservableString.  Part of this work was complete while IB was at the Michigan Center for Theoretical Physics and supported in part by DOE grant DE-FG02-95ER-40899.

\bibliographystyle{utphys}
\bibliography{miracle}

\end{document}